\documentclass[double]{article}
\usepackage{times}
\usepackage{algorithm}
\usepackage{algorithmic}
\usepackage{wrapfig}
\usepackage{verbatim}
\usepackage{amsmath}
\usepackage{graphicx}
\usepackage{subcaption}
\usepackage[active]{srcltx}
\usepackage{color}
\usepackage{lineno}
\usepackage{dirtytalk}
\usepackage{amsthm}
\usepackage{authblk}
\usepackage{listings}

\usepackage{epsfig,graphicx,amsfonts}
\usepackage{amsmath,amsthm,amssymb}
\usepackage{xcolor}

\usepackage{adjustbox}
\usepackage{multirow}
\usepackage{setspace}
\usepackage{hyperref}
\usepackage{comment}

\long\def\comment #1\commentend{}

\begin{document}

\title{\Large Optimizing SMS Reminder Campaigns for Pre- and Post-Diagnosis Cancer Check-Ups using Socio-Demographics: An In-Silco Investigation Into Bladder Cancer}
\author{Elizaveta Savchenko$^{1*}$, Ariel Rosenfeld$^{2}$, Svetlana Bunimovich-Mendrazitsky$^{1}$\\
\(^1\) Department of Mathematics, Ariel University, Ariel, Israel\\
\(^2\)  Department of Information Science, Bar Ilan University, Ramat-Gan, Israel\\
\(*\) Corresponding author: svetlanabu@ariel.ac.il

}

\maketitle 

\date{ }

\begin{abstract}
Timely pre- and post-diagnosis check-ups are critical for cancer patients, across all cancer types, as these often lead to better outcomes. Several socio-demographic properties have been identified as strongly connected with both cancer's clinical dynamics and (indirectly) with different individual check-up behaviors. Unfortunately, existing check-up policies typically consider only the former association explicitly. In this work, we propose a novel framework, accompanied by a high-resolution computer simulation, to investigate and optimize socio-demographic-based SMS reminder campaigns for cancer check-ups. We instantiate our framework and simulation for the case of bladder cancer, the 10th most prevalent cancer today, using extensive real-world data. Our results indicate that optimizing an SMS reminder campaign based solely on simple socio-demographic features can bring about a statistically significant reduction in mortality rate compared to alternative campaigns by up to 5.8\%. \\ \\

\end{abstract}

\maketitle \thispagestyle{empty}

\pagestyle{myheadings} \markboth{Draft:  \today}{Draft:  \today}
\setcounter{page}{1}

\section{Introduction}
\label{sec:introduction}
Cancer is a generic name for a wide range of diseases in which cells in the human body grow and reproduce uncontrollably, resulting in a broad spectrum of clinical conditions and complications, commonly resulting in low life quality and early death \cite{cancer}. In addition to its potentially deadly clinical consequences, cancer is also associated with poor quality of life and a substantial economic burden on patients, their families, and the entire healthcare systems  \cite{cancer_cost_ariel_1,cancer_cost_ariel_2}. The exact causes of cancer are yet to be fully understood, but a combination of genetic and  environmental factors, including socio-demographic ones, are known to be strongly linked with the onset and progression of the disease \cite{cancer_intro_ariel_1}.

There are many different types of cancer and each one may have a different set of risk factors and causes. Nonetheless, it is generally acknowledged that the early detection of the disease (via pre-diagnosis check-ups) and its appropriate monitoring for recurrence (via post-diagnosis check-ups) are pivotal in determining treatment options, reducing treatment costs, improving quality of life, and, arguably most important, lowering mortality rates across all patient groups \cite{intro_1,intro_2,intro_3,intro_4}. Thus, developing and implementing proper cancer check-up policies, both Pre- and Post-Diagnosis (PPD), is crucial \cite{intro_5}. 
Unfortunately, determining an optimal PPD check-up policy for a given individual is still an open, yet active, area of research \cite{rw_review_1,rw_review_2,journal_1,journal_2}. For example, \cite{rw_1} reviewed multiple policies for breast cancer PPD check-ups and found that the current policies lack clinical or economic evidence as to their effectiveness from a healthcare service provider (HSP) perspective. 
In a similar manner, \cite{rw_3} performed an evaluation of 20 check-up policies for breast cancer by considering the various costs associated with implementing these policies and the potential subsequent medical costs. The authors found that policies that are specifically tailored to different age groups result in significantly better outcomes. Accordingly, in order to derive socio-demographic-based PPD check-up policies, researchers commonly use mathematical models, a practice which has proved to be very powerful  \cite{math_good_1,math_good_2,math_good_3,math_good_4,math_good_5}. That is, researchers rely on data-driven models, usually trained with machine learning algorithms, and optimization techniques to derive approximated or optimal PPD check-up policies in a fast, secure, and affordable manner \cite{math_good_cancer_1,math_good_cancer_2,math_good_cancer_3,math_good_cancer_5}. 

Unfortunately, an optimized PPD check-up policy need not necessarily be followed by all individuals alike  \cite{ppd_visit_1,ppd_visit_2}. That is, the real-world effectiveness of a PPD check-up policy strongly depends on individual compliance \cite{latest} which, in turn, is known to be strongly linked to one's socio-demographic characteristics \cite{ppd_more_people_2}.
For example, \cite{example_socio_intro} showed compliance to colorectal cancer screening is significantly higher in women than in men and changes non-linearly with age. In order to increase individual compliance, particularly in high-risk patient groups, various stakeholders such as HSPs and governmental agencies have been implementing diverse compliance-increasing strategies such as health education programs, taxation, discount offers, and SMS reminder campaigns \cite{ppd_more_people_1,ppd_more_people_3}. These strategies differ in their effectiveness, costs, and operational overhead. However, SMS reminder campaigns are often considered to be very effective, cheap, flexible, and operationally simple to implement compared to the mentioned alternatives. For example, \cite{sms_policy_1} reviewed seven research projects concerning SMS reminder campaigns in Africa, concluding that vaccination reminder has led to improvements in vaccination uptakes under various metrics, whether through the increase in vaccination coverage, decrease in dropout rates, increase in completion rate or decrease in delay for vaccination. In particular, \cite{sms_support} showed that the SMS campaign gain similar clinical benefits to other approaches such as home visits while being significantly cheaper and much more scalable. \cite{sms_policy_2} shown that SMS reminders can be used to reduce health and social inequity while providing better clinical outcomes for patients suffering from Human Immunodeficiency Virus.
Unfortunately, to the best of our knowledge, existing cancer PPD check-up policies are currently accompanied by a na\"{i}ve \say{one-size-fits-all} SMS reminder campaigns, where all patients are treated the same. 

\begin{figure}[!ht]
    \centering
    \includegraphics[width=0.99\textwidth]{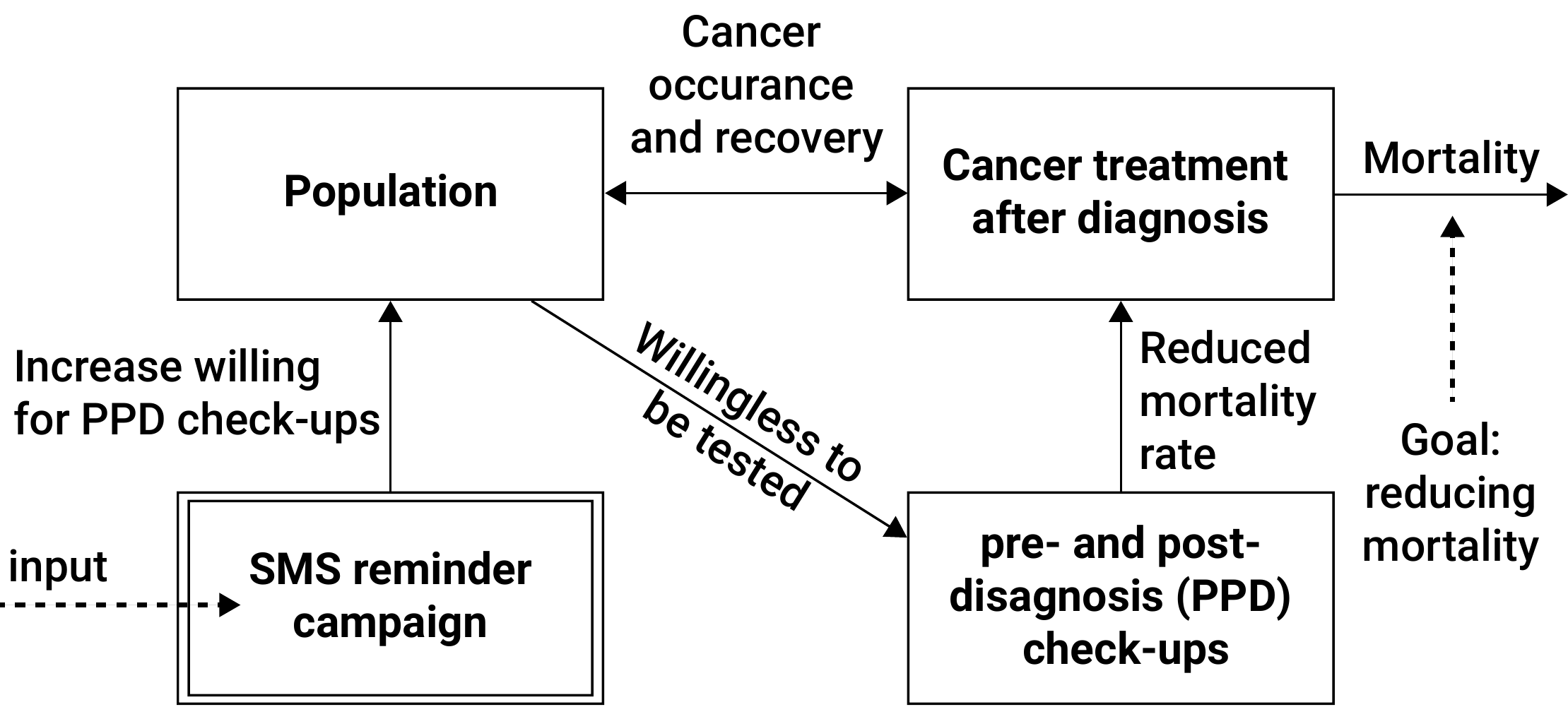}
    \caption{A schematic view of the framework's input and objective and how it interacts with the current components in the dynamic.}
    \label{fig:intro}
\end{figure}

In this work, we propose a novel framework, accompanied by a high-resolution computer simulation, to investigate and optimize a socio-demographic-based SMS reminder campaign for cancer PPD check-ups. Our framework can be instantiated to any type of cancer and PPD policy, for which an optimal socio-demographic-based SMS reminder campaign is approximated through a Monte Carlo optimization technique. Fig. \ref{fig:intro} shows a schematic view of the proposed model's structure, input, and objective. 

We instantiate our framework and provide an in-depth \textit{in silco} investigation into Bladder Cancer (BC). BC is the 10th most common cancer worldwide, with more than half a million new cases yearly and 200 thousand associated deaths in 2018 \cite{Jemal}. A more recent report reveals 34 thousand BC-related deaths during 2021 and 90 thousand new cases in the United States alone \cite{bc_stat_2021}, indicating a growing trend in both metrics. Similar trends are found in many other types of cancer as well \cite{bc_stat_2021}. BC is also associated has a high recurrence rate, invasive surveillance strategies, and high treatment costs which combine to make it the single most expensive cancer to manage in both England and the United States \cite{EYLERT}. As such, BC is a prime candidate for the implementation of compliance-increasing strategies such as SMS reminder campaigns.

The rest of the paper is organized as follows: Section \ref{sec:model} formally presents the framework, followed by Section \ref{sec:results} which outlines its implementation to the case of BC. Finally, in section \ref{sec:discussion}, we analyze and discuss the results as well as propose possible future work directions.  

\section{Framework}
\label{sec:model}
Our proposed framework consists of several interconnected components which are detailed and discussed below. First, we define the clinical dynamics of cancer's onset and progression in the context of PPD check-ups and treatment. Then, we formalize the challenge of determining a PPD check-up policy as a resource-bounded optimization task. Based on these two components, we formulate the SMS reminder campaign optimization task and propose a Monte Carlo optimization technique that is shown to coverage to a near-optimal solution given enough computational resources. Then, we propose a fitting procedure to set the parameters of an instance of the framework using historical data and facilitates the fitting of unavailable parameters' values that agree with realistic scenarios. Last, we detail how the different components are assembled together into a single framework.

\subsection{Clinical dynamics}
\label{sec:clinical_dynamics}
\begin{figure}[!ht]
    \centering
    \includegraphics[width=0.99\textwidth]{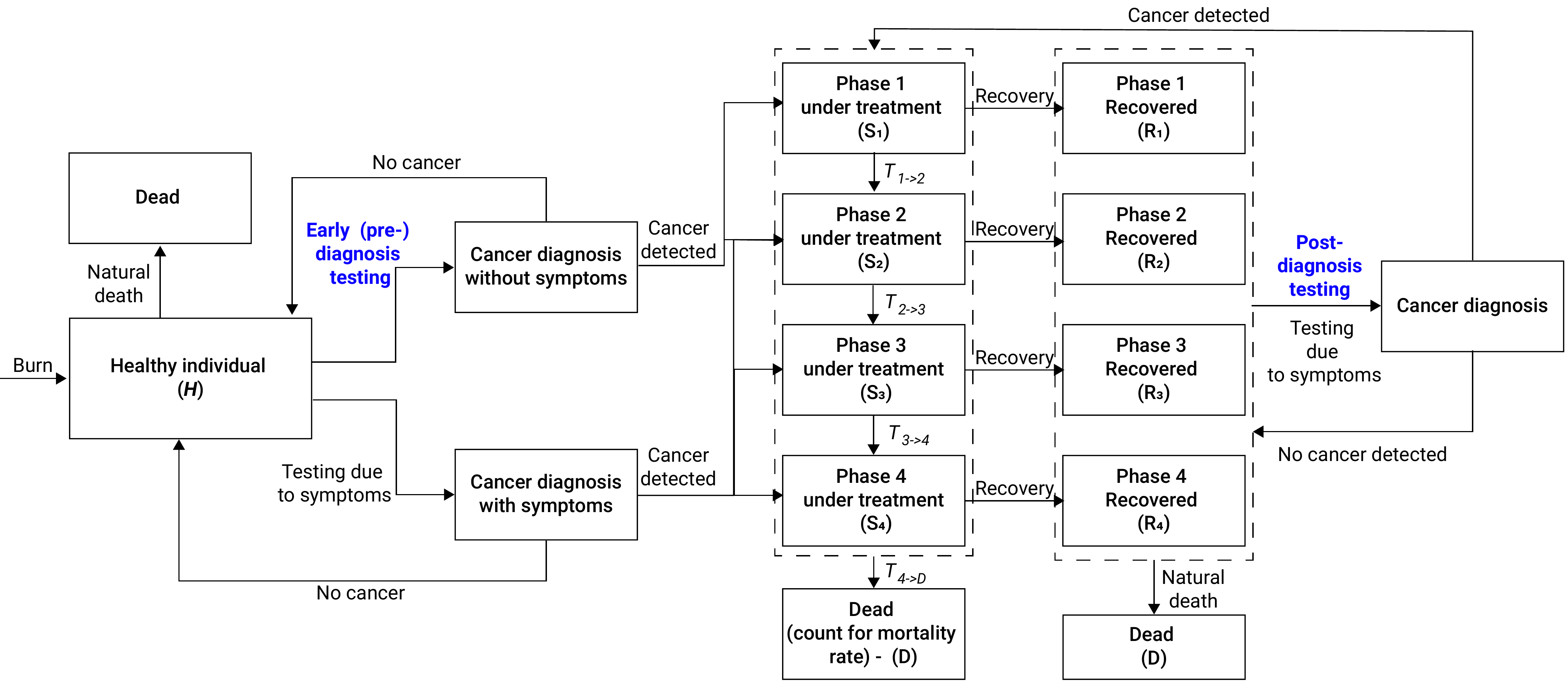}
    \caption{A schematic view of the clinical statuses and the transitions between them due to PPD check-ups and clinical deterioration. The optimization components of the framework are highlighted in blue.}
    \label{fig:individual_flow}
\end{figure}

Individuals are categorized into one of 10 clinical-oncological statuses (denoted by their \(\alpha\) parameter): healthy \((H\)), sick at phase \(j\) (\(S_j\) such that \(j \in \{1, 2, 3, 4\}\)) Recovered from phase \(j\) (\(R_j\) such that \(j \in \{1, 2, 3, 4\}\)), and dead \((D)\) such that \(N = H + S_1 + S_2 + S_3 + S_4 + R_1 + R_2 + R_3 + R_4 + D\) where \(N\) is the population's size at a given point in time. Individuals in the first (healthy) status were never diagnosed with cancer (\(H\)). If the individual never gets sick with cancer, it eventually naturally dies after \(\gamma\) steps in time and transforms to the dead (\(D\)) status. Healthy individuals can perform a pre-diagnosis check-up by either following the existing pre-diagnosis policy or due to symptoms. We assume that if the individual suffers from symptoms, s/he will choose to perform a pre-diagnosis check-up regardless of the PPD policy. A policy-based pre-diagnosis check-up will result in one of three outcomes: either indicating that the individual is healthy (\(H\)) or s/he has cancer of phase 1 or 2 (\(S_1, S_2\)). Note that cancer of phase 3 must include significant symptoms that are assumed to be noticeable by the medically-unprofessional patient such as extreme pain. In a similar manner, a check-up due to symptoms will either result in a non-cancer diagnosis (i.e., healthy (\(H\)) from a cancer perspective -- potentially indicating a non-oncological disease), or a cancer diagnosis with either phase 2, 3, or 4 (\(S_2, S_3, S_4\)). Once an individual is diagnosed, treatment takes place immediately. Each individual has a personal duration and probability to recover, according to their socio-demographic properties and cancer phase. If the individual dies during the treatment, it transforms into the dead (\(D\)) status, and mortality due to the disease is recorded. Otherwise, the individual recovers and transforms to the corresponding recovery phase \((S_j \rightarrow R_j)\). Similar to healthy individuals, any other individual eventually naturally dies after \(\gamma\) steps in time and transforms into the dead (\(D\)) status. Here, we assume that getting sick with cancer does not affect the individual's life expectancy if s/he recovers from it. Similar to healthy individuals, recovered individuals can perform post-diagnosis check-ups following the post-diagnosis policy and/or due to symptoms. Here, both check-up types may result in a healthy outcome (i.e., the individual remains in the same clinical status) or in a cancer outcome (i.e., recurrent cancer) of phase \(k \in [1, 2, 3, 4]\) \cite{re_cancer_1,re_cancer_2}. During the illness, the transition from one phase to the consecutive one, until death, follows a socio-demographic-based dynamics indicated by \(T_{1 \rightarrow 2}\), \(T_{2 \rightarrow 3}\), \(T_{3 \rightarrow 4}\), and \(T_{4 \rightarrow D}\). Fig. \ref{fig:individual_flow} provides a schematic view of the clinical statuses and the flow between them as a result of PPD check-ups. 

\subsection{PPD check-up policy}
Each individual in the population \(p \in P\) is represented by a timed finite state machine \cite{fsm} as follows: \(p := (\alpha, \tau, \mu, \rho, e, g, s, \gamma)\) where \(\alpha \in \{H, S_1, S_2, S_3, S_4, R_1, R_2, R_3, R_4, D\}\) is the current clinical status of the individual, \(\tau \in \mathbb{N}\) is the time passed from the last change of the clinical status \((\alpha)\), \(\mu \in [0, 1]\) is the probability that the individual will naturally comply with the pre-diagnosis check-up (without any compliance-increasing strategy), \(\rho \in \mathbb{R}^+\) is  the individual's degree of openness or susceptibility to compliance-increasing strategies, \(e \in \mathbb{N}\) is the individual's age, \(g \in \{male, \; female\}\) is the individual's gender, \(s \in [1, 2, \dots, 10]\) is the relative socio-economic tenth percentile of the individual, and \(\gamma \in \mathbb{N}\) the time steps until the individual naturally dies. 

From a socio-demographic perspective, the model uses \(12\) parameters as indicated by the tuple \((e, g, s)\): \(T_{1 \rightarrow 2}, T_{2 \rightarrow 3}, T_{3 \rightarrow 4}, T_{4 \rightarrow D}, b, \gamma, \psi_i, \psi_r^j, \delta_i, \delta_r\) where \(\psi_i \in [0, 1]\) is the probability that an individual would get sick with cancer for the first time, \(\psi_r^j \in [0, 1]\) is the probability that the individual would get sick with recurrent cancer after recovering from phase \(j \in \{1, 2, 3, 4\}\), \(\delta_i\) is the duration delta between two consecutive pre-diagnosis check-ups recommended to that individual, and \(\delta_r\) is the duration delta between two consecutive post-diagnosis check-ups recommended to that individual. We assume that the life expectancy of an individual, \(\gamma\),  is dependent on the socio-economic status (\(s\)) and gender (\(g\)) alone. 

It is important to note that we consider PPD check-up policies to be mere \textit{recommendations} that cannot be enforced on any individual.

\subsection{SMS reminder campaign}
\label{sec:policy_optimization}
SMS (Short Message Service) reminder campaigns are a popular compliance-increasing strategy in healthcare \cite{sms_policy_1,sms_policy_2}. The implementing agency (e.g., HSP) can send any number of reminders to any subset of patients in order to encourage them to comply with the PPD policy \cite{sms_repeat}. Formally, an SMS reminder campaign is a function, \(\Phi\), that accepts the population, represented by a set of finite state machines, over time and returns when and how many SMSs each socio-demographic group (defined by the \(e, g,\) and \(s\) parameters of each individual) should get. Each SMS increases an individual's likelihood to follow the PPD policy as proposed by \cite{sms_policy_1}:
\[
    \mu \leftarrow \mu + \rho \Big ( c_1 + c_2 \big ( log_{10} (n) - log_{10} (n-1) \big ) \Big ), 
\]
where \(c_1\) and \(c_2\) are the SMS effectiveness coefficients and \(n\) is the number of SMSs the individual received thus far. In addition, each SMS has a fixed cost \(b \in \mathbb{R}^+\). Since the implementing agency is limited by some budget \(B \in \mathbb{R}^+\) for a fixed duration \([t_0, t_f]\), deriving an SMS reminder campaign can be formulated as the following resource-bounded optimization task:
\begin{equation}
    \min_{\Phi} MR_{[t_0, t_f]}(\Phi) \; \text{s.t.} \; cost(\Phi) \leq B,
    \label{eq:policy_opt_task}
\end{equation}
where \(MR_{[t_0, t_f]}\) is a function that returns the average mortality rate during  \([t_0, t_f]\) and \(cost(\Phi)\) is a function that returns the total cost of an SMS reminder campaign \(\Phi\). 
Note that \(\Phi\) makes decisions in a discrete manner. Since \(\Phi\) is not able to pick a specific individual from each socio-demographic group as the \(\mu\) and \(\rho\) parameters of each individual are not available to \(\Phi\) in realistic cases, an SMS is sent in random inside each socio-demographic group in an equally distributed manner. 

In order to solve the SMS reminder campaign optimization task (Eq. (\ref{eq:policy_opt_task})), we used a Monte Carlo approach \cite{optimization}. Namely, we sample the SMS reminder campaign parameter space. It takes a form of a four-dimensional tensor with one temporal dimension and three dimensions representing the age, gender, and socio-economic status of each socio-demographic group. Each value in the resulting matrix represents the relative part of the entire budget (\(B\)) allocated to SMS distribution among each socio-demographic group at each step in time. After the parameter values are set, we run the model for \(t_f - t_0\) rounds and calculate the average mortality rate. If a configuration resulted in an average mortality rate smaller than any previous parameter configuration, we declare this configuration to be the best one so far. Since the parameter configuration space is finite, this computational procedure is guaranteed to converge to the optimal solution as the number of samples goes to infinity \cite{mc_good}. Overall, for the case of stochastic processes with random functions that are piecewise convex and a discrete state space, such as the case here,  it has been proven elsewhere that this optimization process has a probability to reach the optimal solution that approaches one exponentially fast with an increase in the number of simulations \cite{mc_converage_rate}.

\subsection{Fitting procedure}
\label{sec:histrocial_hit}
In order to obtain the parameters that best fit historical records, we use the gradient descent (GD) method for the parameters' space following \cite{historical_fit}. Formally, given the model's initial condition, the parameter space, historical data, and a loss function \(d\) we use the GD method \cite{gradient_descent_method} to find the parameters that minimize \(d\) on a fixed and finite duration in time \([t_0, t_f]\) such that \(t_0 < t_f\). Formally, let us denote the parameter space   by \(\mathbb{P} \in \mathbb{R}^\epsilon\) where \(\epsilon \in \mathbb{N}\) is the number of parameters in the implemented framework. In addition, a specific parameter configuration is denoted by \(P \in \mathbb{P}\). We also denote the parameter configuration of the \(i_{th}\) iteration of the GD by \(P_i\). Since the GD is computed on the parameter space with respect to a loss function \(d\), the gradient is numerically obtained by following the five-point stencil numerical scheme \cite{numerical_scheme}:
\[
    \nabla P_i := \forall j \in [1, \dots \epsilon]: \frac{-d(P_i(j, -2)) + 8d(P_i(j, 1)) - 8d(P_i(j, -1)) + d(P_i(j, 2))}{12h}
\]
where \(P_i(j, k) := P_i(x_1, \dots, x_j + kh, \dots, x_\epsilon)\) and \(h \in \mathbb{R}^+\) is the step's size. 
\[
    P_{i+1} \leftarrow P_{i} + \nabla P_i.
\]
For our case, let us assume an instance of the framework, \(M_P\), with the  parameter configuration \(P\). In order to fit on historical data, we define a metric \(d\) between the prediction of the mortality rate and the historically recorded mortality rate:
\begin{equation}
  \begin{array}{l}
\left.\begin{aligned}
    d(M_C, H) := \Sigma_{t=t_0}^{t_f} |MR(M_C, t) - H(t)|,
\label{eq:dynamics_metric}
\end{aligned}\right.
  \end{array}
\end{equation}
where \(H\) is the historical recorded mortality rate, \(MR\) is a function that gets the simulation's state as defined by the distribution of the individuals' clinical state \(M_P\), and \(t \in \mathbb{N}\) is a point in time and returns the average predicted mortality rate of the framework at this point in time. We define \(d\) to be the mean absolute error of the predicted mortality rate since this is the metric that one, presumably, wishes to minimize using a designated SMS reminder campaign. 

\subsection{Assembling the components into a single framework}

\begin{figure}[!ht]
    \centering
    \includegraphics[width=0.75\textwidth]{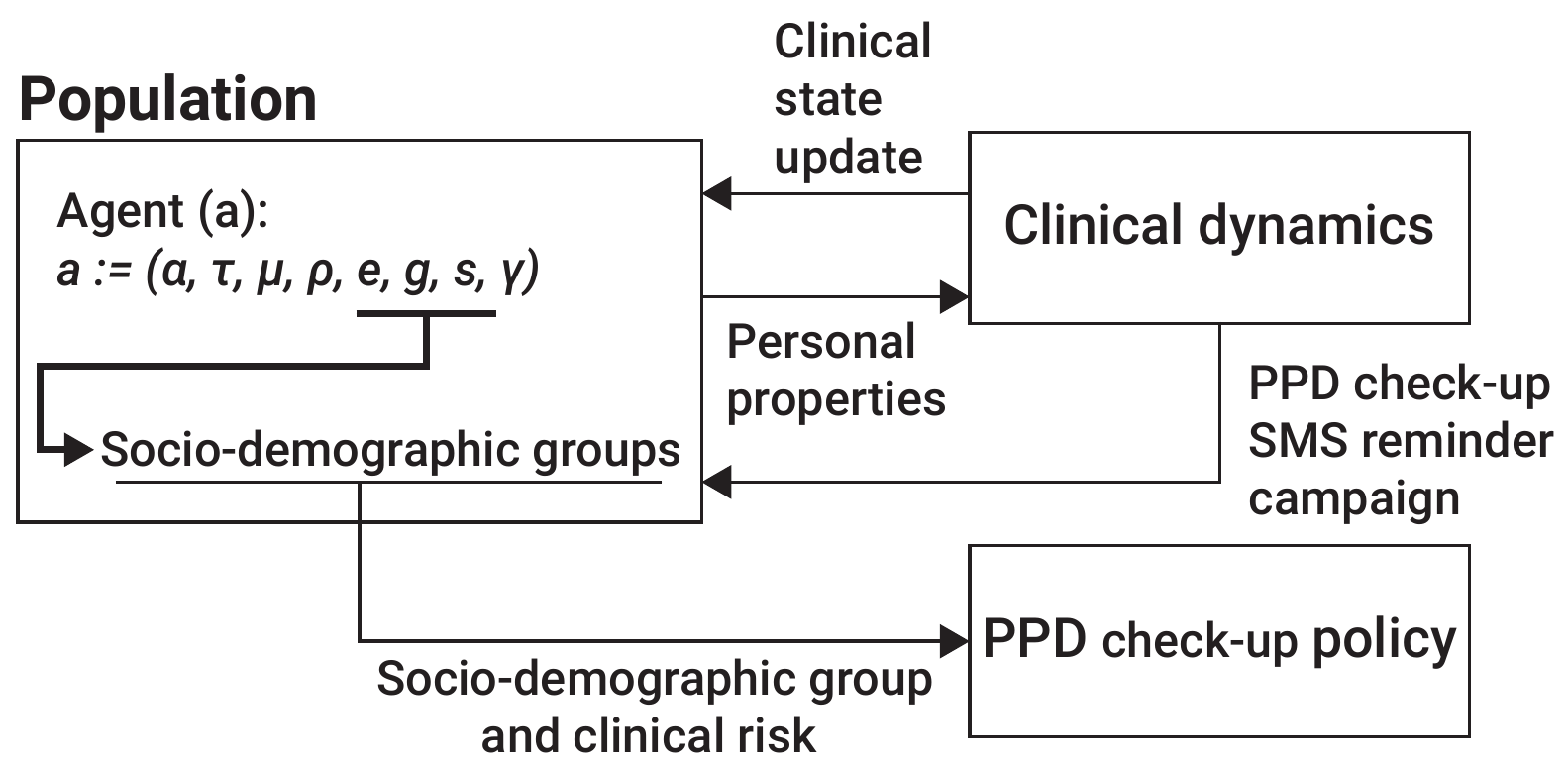}
    \caption{A schematic view of the framework's components and the interactions between them. Recall that the socio-demographic groups are defined by the individual's age (\(e\)), gender (\(g\)), and socio-economic status (\(s\)). These properties, along with clinical risk, determine the PPD check-up policy and the accompanying SMS reminder campaign. All personal features together are reflected in the clinical dynamics.}
    \label{fig:ppd}
\end{figure}

The framework has a global and discrete clock that all the individuals in the population follow. Namely, let us define each step in time as a round \(t \in [1, \dots, T]\), where \(T<\infty\). In the first round (\(t=1\)), the population is created to satisfy a pre-define co-distribution of socio-demographic and clinical distributions. The PPD check-up policy is defined and fixed at this stage as well. Then, at each round \(t \geq 1\), each individual is following the PPD policy with its personal probability \(\mu\), according to its clinical status \(\alpha\). Afterward, the clinical dynamics are executed for each individual in the population, in a random order. Right after, the SMS reminder campaign is activated based on the population's state, sending, if any, SMSs to each socio-demographic group according to \(\Psi\). In addition, the population naturally grows at a rate, \(r \in \mathbb{R}\) which corresponds to its current size. All born individuals assume to keep the population's socio-demographic co-distribution as identical as possible and set to be healthy (\(\alpha \leftarrow H\)). Fig. \ref{fig:ppd} shows a schematic view of the framework's components and the interactions between them. 

\section{Investigation Into Bladder Cancer}
\label{sec:results}
The following analysis consists of three parts: First, we outline the implementation of the above generic framework for the case of BC in the United States. Then, we propose five candidate SMS reminder campaigns targeted at minimizing the expected mortality rate. Last, we explore the characteristics of the best-performing SMS reminder campaign and its statistical relationship to its underlying socio-demographic characteristics.

\subsection{BC Implementation}
\label{sec:setup}
For realising the proposed framework, several parameters have to be set. Since most of the relevant data is available for the United States, it is the focus of our analysis. We rely on the following sources \cite{data_1,data_2,data_3,data_4,data_5,data_6,data_7,data_8} which are accrued, integrated, pre-processed for our needs and made available as a data file in the supplementary material. 
Specifically, in order to obtain the population's growth rate (\(b\)), we use the data of the United States population's growth between 1950 and 2020\footnote{\url{https://ourworldindata.org/grapher/population-growth-rate-with-and-without-migration}}, fitting a standard exponential smoothing time-series forecasting model \cite{exp_forcast}. The window size for this forecasting model is obtained using the grid search method, ranging between 2 and 25 years \cite{grid_search} and aiming to minimize the mean absolute error.
For the life expectancy, \(\gamma\), we use the United States average life expectancy as reported by the United Nations\footnote{\url{https://www.macrotrends.net/countries/USA/united-states/life-expectancy}}, ranging between 1950 and 2020. In order to get the life expectancy divided into gender and socioeconomic status, we used life expectancy gender differences reported by \cite{le_gender_diff} and the socioeconomic status differences reported by \cite{le_economic_status_diff}. When the socioeconomic status is divided by tenths, this division results in 20 time-series  with 12 constraints - 10 for the socioeconomic status, one for the gender differences, and one to agree with the average reported life expectancy. Since there are more relationships than constraints, there is an infinite number of possible solutions to this computational task. We find a feasible solution using the least mean square method \cite{leastSquares}, obtaining a time-series function for the life expectancy for each gender and socioeconomic status separately in the same way the population's growth rate was obtained.

As the underlying PPD check-up policy, we consider the recommendation of the American Cancer Society\footnote{\url{https://www.cancer.org/}}. Specifically, individuals are encouraged to perform a pre-diagnosis check-up once a year following the age of 45 (none before that) and post-diagnosis check-ups once a year for those recovered from phase 1 or 2, or twice a year for those recovered from phase 3 or 4 \textit{regardless of age}. That is, people of all ages may be encouraged to perform check-ups and thus get SMS reminders. In order to find the SMS effectiveness coefficients (\(c_1, c_2\)), we used the data reported by \cite{sms_policy_1} and fitted it using the least mean square method \cite{leastSquares}. In addition, we averaged the SMS sending cost of five leading SMS providers in the US, as manually sampled in 2023, obtaining an average SMS cost of 0.049 US dollars. 

For initialization, we used the age, gender, and socioeconomic data from the US in 2022 as reported by the US Census Bureau\footnote{\url{https://www.census.gov/topics/population/age-and-sex/data/tables.html}}. Since the data is not provided as the cross of all three properties (i.e., the number of individuals for each combination of age, gender, and socioeconomic status), we assume the age and gender distributions for each socioeconomic group are identical. This assumption is known to be false but it is commonly adopted due to a lack of finer-grained data at a publicly available level \cite{gender_corr_money}. Overall, we include 333 million individuals in the initial condition, divided into 140 socio-demographic groups (i.e., two gender groups, seven age groups, and ten socioeconomic groups). In addition, we assume a budget of ten million dollars. We set \(t_0 = 0\) and \(t_f = 365\) and each round, \(t_i\), to be a single day.  

\subsection{Parameter fitting}
\begin{figure}[!ht]
    \centering
    \includegraphics[width=0.99\textwidth]{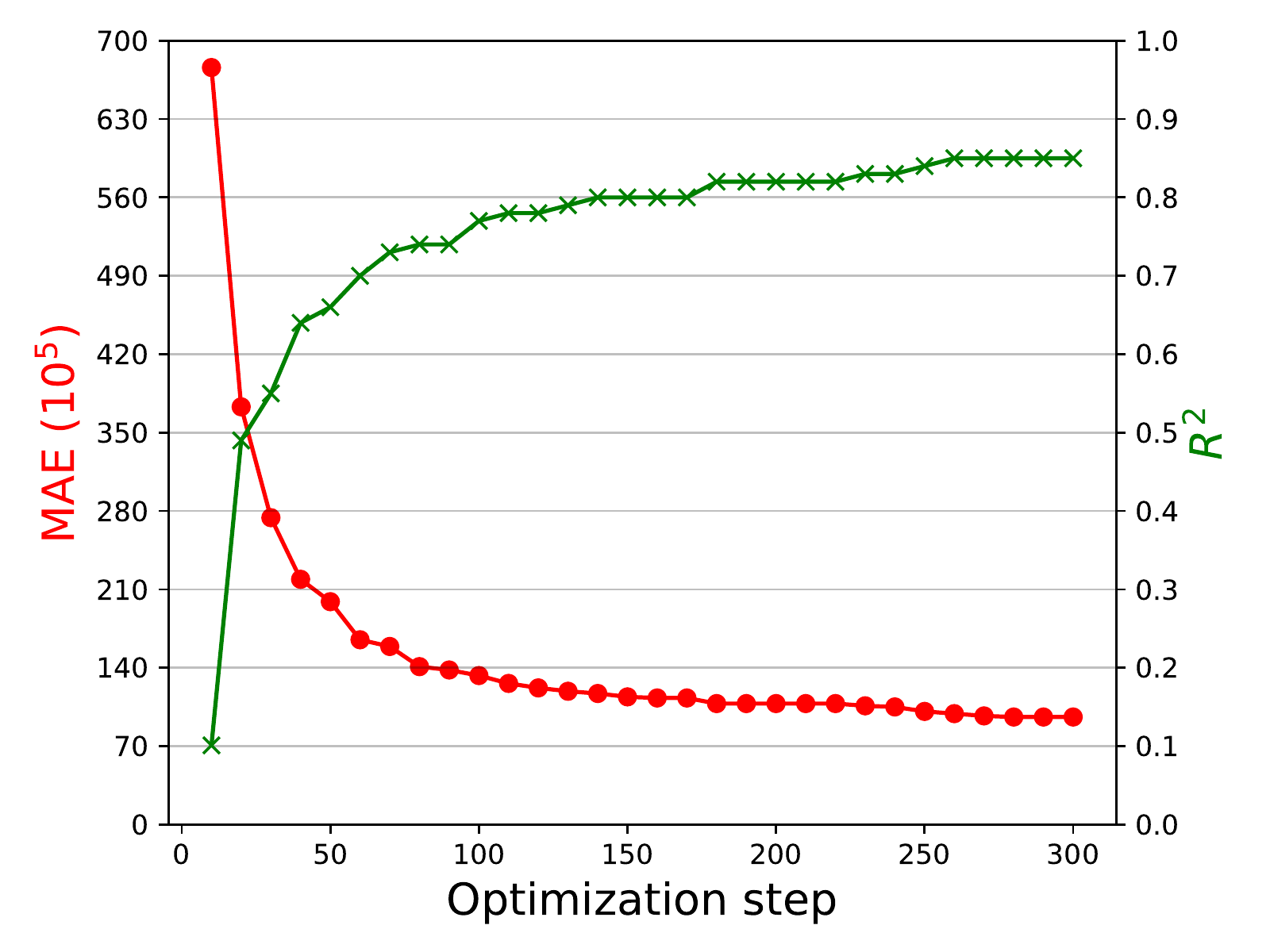}
    \caption{The fitting process. The MAE (red line) decreases as a function of the simulation steps while the \(R^2\) (green line) is increasing. The results are shown as the mean of \(n=50\) simulations.}
    \label{fig:historical_fit}
\end{figure}

Given the framework's initial condition and available parameter values (see Section \ref{sec:setup}), we used the proposed fitting  procedure (see Section \ref{sec:histrocial_hit}) in order to obtain the remaining parameter values that best align with historical data. Notably, as part of this process, we find the values of \(\mu\) and \(\rho\) across the population. This is important as these values are not readily available and can only be approximated by fitting a model that describes the dynamics of the observed historical data. Fig \ref{fig:historical_fit} shows the model's MAE from the historical data (red line), as defined by Eq. (\ref{eq:dynamics_metric}), and the coefficient of determination (\(R^2\)) (green line). Due to the stochastic nature of the framework, the results are shown as the mean of \(n=50\) simulations. We fitted both signals using the SciMed symbolic regression model \cite{scimed}, obtaining: \(MAE(i) = 329i^{-0.384}\) and \(R^2(i) = 0.517 + 0.104ln(i)\) with \(R^2 = 0.969\) and \(R^2 = 0.967\), respectively, where \(i \in [10, 20, \dots, 290, 300]\) is the optimization step's index. One can notice that the optimization process coverage around an MAE of \(9.5 \cdot 10^7\) with an \(R^2\) of \(0.85\). 

\subsection{SMS reminder campaigns}
We compare five SMS reminder campaigns with increasing levels of sophistication: \say{None}, Na\"{i}ve, Greedy, Na\"{i}ve Monte Carlo, and Socio-demographic Monte Carlo. The first, the \say{None} campaign, indicates that there are no SMS reminders sent at all. The Na\"{i}ve campaign treats all individuals as if they belong to the same (single) socio-demographic group. As such, the Na\"{i}ve campaign treats individuals differently only based on the PPD check-up policy and their clinical state (\(\alpha\)). 
The Greedy campaign takes into consideration the mortality rate associated with each socio-demographic group such that the allocated budget to each group is proportional to its relative contribution to the overall mortality rate in the entire population. Namely, the Greedy campaign optimizes the SMS reminder campaign in every single step in time, ignoring the need to optimize for the entire duration \([t_0, t_f]\). Finally, we evaluate two variants of the Monte Carlo optimized SMS reminder campaigns - a Na\"{i}ve one, which does not consider the socio-demographic characteristics of each individual and thus it only optimizes for the timing and volume of reminders sent to each individual based on their clinical statuses (like the Na\"{i}ve campaign); and a socio-demographic one which applies the socio-demographic-based Monte Carlo optimization technique (as formally described in Section \ref{sec:policy_optimization}).
Both variants were trained for 10,000 instances before being applied in the following analysis.

\begin{figure}[!ht]
    \centering
    \includegraphics[width=0.99\textwidth]{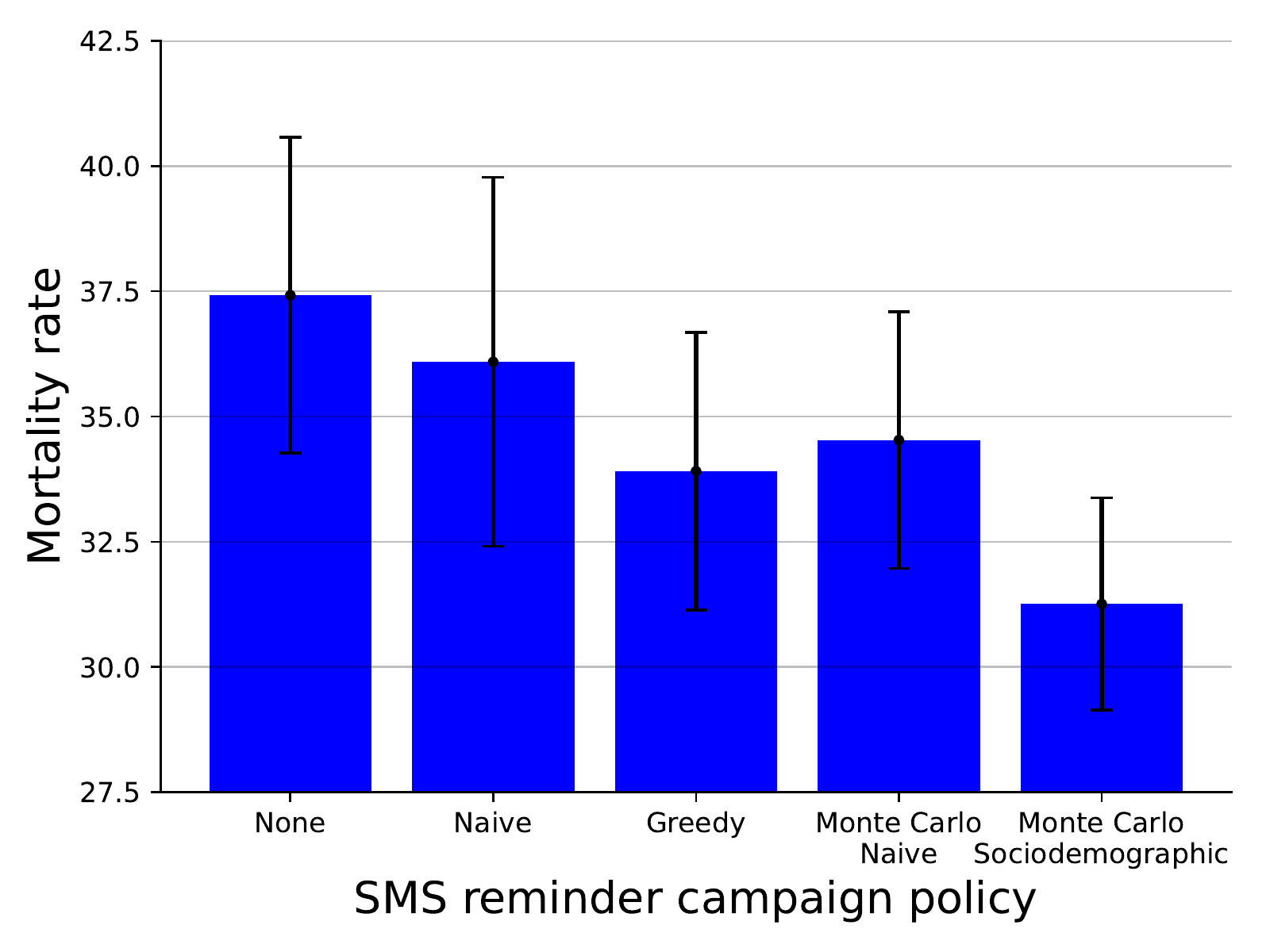}
    \caption{Average mortality rate across the five examined SMS reminder campaigns. The results are shown as the mean \(\pm\) standard deviation of \(n=100\) simulations.}
    \label{fig:policies}
\end{figure}

Fig. \ref{fig:policies} presents the comparison of these five SMS reminder campaigns. The results are shown as the mean \(\pm\) standard deviation of \(n=100\) simulations. One can notice that the Monte Carlo socio-demographic campaign (the right-most column) results in the lowest average mortality rate. In order to validate this outcome statistically, we computed ANOVA test \cite{anova} with a one-sided T-test post-hoc correction \cite{t_test}, obtaining that, indeed, this campaign is statistically better compared to the alternative campaigns with \(p < 0.01\).

\subsection{Campaign analysis}
\begin{figure}[!ht]
    \begin{subfigure}{.5\textwidth}
        \includegraphics[width=0.99\textwidth]{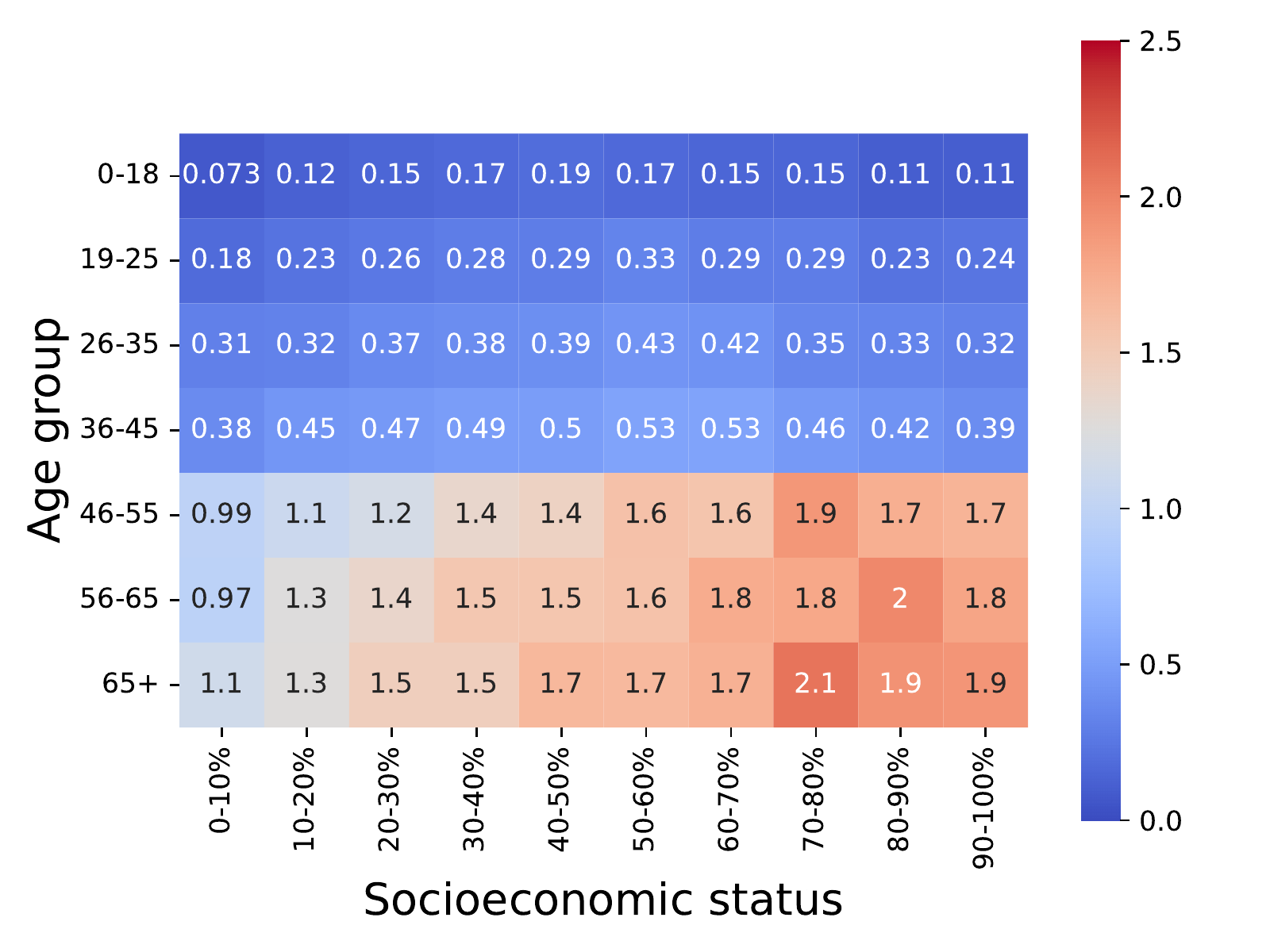}
        \caption{Male.}
        \label{fig:males}
    \end{subfigure}
    \begin{subfigure}{.5\textwidth}
        \includegraphics[width=0.99\textwidth]{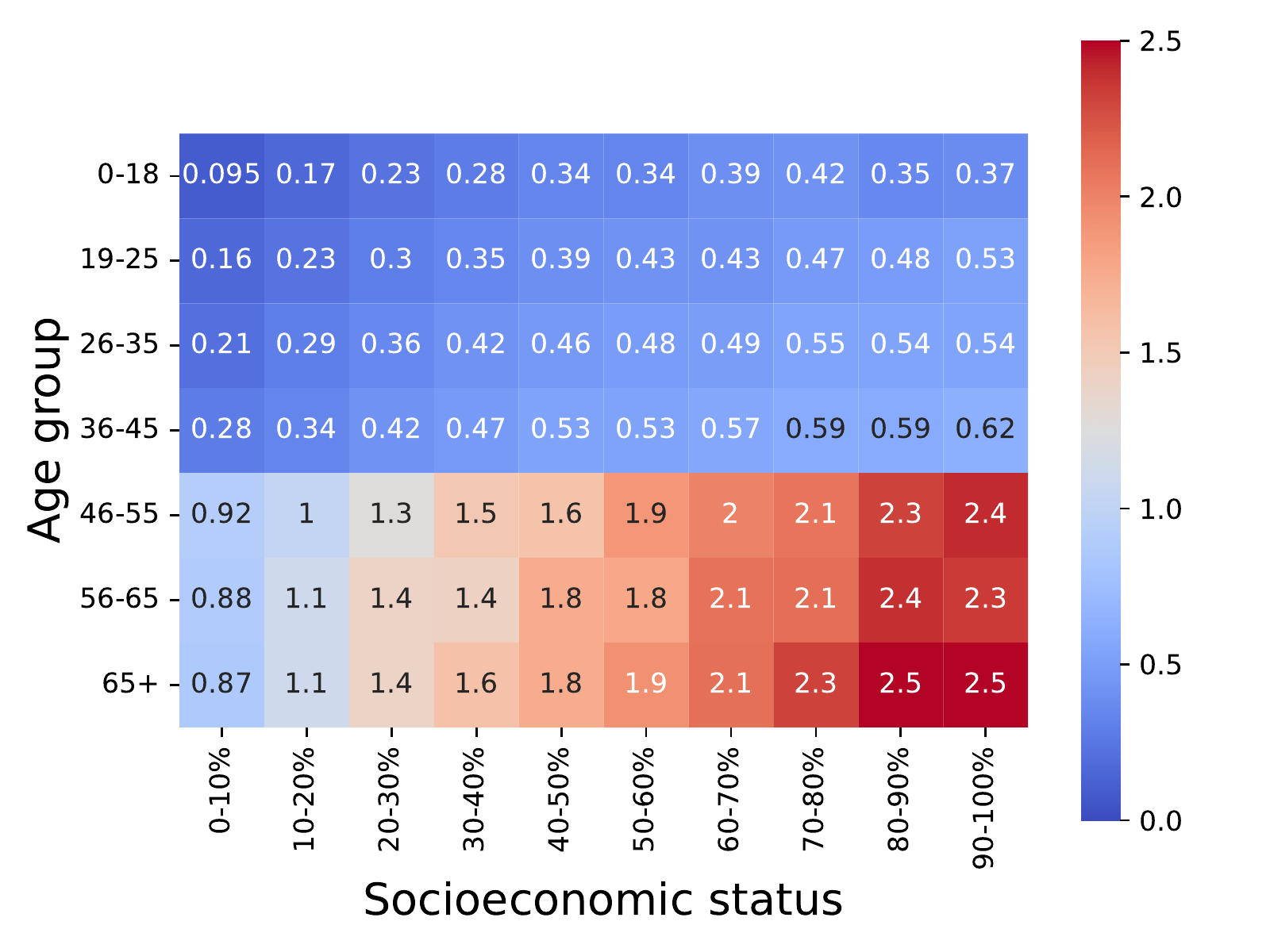}
        \caption{Female.}
        \label{fig:real_wolrd_process_2}
    \end{subfigure}
    
    \caption{The average yearly number of SMSs an individual in each socio-demographic group would obtain based on the Monte Carlo socio-demographic SMS reminder campaign.}
    \label{fig:analysis}
\end{figure}

We further explore the best-performing SMS reminder campaign and analyze the relative amount of resources invested in each socio-demographic group. Fig. \ref{fig:analysis} shows the average yearly number of SMSs an individual in each socio-demographic group would get based on the Monte Carlo socio-demographic SMS reminder campaign, divided into male and female heatmaps. The figure shows that, generally speaking, older individuals with higher socio-economic status require more resources. In particular, a sharp shift in resource allocation is observed around age 45 and the 40\% percentile of socio-economic statuses. In addition, for most of the age and socio-economic statuses, females get more reminders. Similar results were obtained when testing for two, five, fifteen, and eighteen million dollars budget with less than a five percent difference in the relative number of SMS an individual in each socio-demographic group gets between the two and eighteen million dollars budget.

\section{Discussion}
\label{sec:discussion}
In this study, we proposed a novel framework and simulation that allows for the optimization and investigation of SMS reminder campaigns for cancer check-ups. Considering BC in the US as a representative example, we implemented the framework based on real-world historical data, and derived a Monte Carlo optimized SMS reminder campaign based on individuals' socio-demographic characteristics. The resulting campaign is shown to favorably compare with sensible alternatives.

Our results first demonstrate the presumed adequacy of our proposed framework, at least in the case of BC. As can observe in Fig. \ref{fig:historical_fit}, after fitting on historical data, the implemented framework can explain up to \(85\%\) of the BC-related mortality variance. Using the currently practiced PPD check-up policy, a near-optimal socio-dempogrhaic SMS reminder campaign is found to be superior to several alternatives as presented in Fig. \ref{fig:policies}. 
This result generally agrees with the previous research on this subject \cite{ppd_visit_1,ppd_visit_2}. Namely, taking socio-demographic data into consideration leads to better performance of an SMS reminder campaign. Moreover, campaigns that operate for a long duration and consider the population's clinical distribution seem to operate more favorably compared to greedy approaches which utilize the SMS reminder campaign's budget quickly, raising much awareness yet in a short amount of time. 

The derived campaign was also analyzed to identify the more targeted sub-populations. Fig. \ref{fig:analysis} suggests that the resource distribution is biased towards females and older individuals with a higher socioeconomic status. First, a sharp increase in the average yearly number of SMSs an individual gets occurred crossing the age of 45 years old. This shift can be easily explained by the practiced PPD check-up policy which indicates that pre-diagnosis check-ups are recommended only for individuals older than 45 years old. As such, up to this age group, the SMS reminder campaign is only required for post-diagnosis testing reminders. In addition, one can notice that the number of SMSs increases with age, as the number of cancer-recovered individuals is also increasing with age following more new cases as well as more individuals that recovered and stay healthy since. The slight increase towards higher socio-economic statuses can be associated with the \(\mu\) and \(\rho\) parameters of these subpopulations. Namely, individuals of higher socio-economic status may be associated with higher alternative costs to come for a check-up and therefore require more SMSs on average, as also suggested by \cite{rw_4}. 
Taken jointly, our results suggest that socio-demographic-aware SMS reminder campaigns for cancer PPD check-ups could prove extremely valuable even under strict budget constraints. 

This study has important limitations which offer fruitful avenues for future research. First, it is assumed that the PPD check-ups are perfect and produce 100\% accurate and reliable results as to an individual's clinical status. Unfortunately, this is not true for most clinical tests in general, and oncological tests in particular \cite{test_accuracy}. Thus, our presented results should be treated as slightly over-optimistic of the expected realistic outcomes. In future work, we intend to tackle this shortcoming by integrating PPD check-up accuracy data. Second, alternative optimization techniques could be implemented to derive optimal SMS campaigns or potentially reduce the computational burden of obtaining a near-optimal one. Similarly, the lack of self-explainability of the resulting campaign should be tackled in order to promote its acceptance by stakeholders \cite{sr_2,sr_3,liza,sr_1,latest_1,latest_2}. Finally, when considering BC specifically, an individual's occupation, smoking habits, and other contextual characteristics are known to be strong indicators for the risk of developing the disease \cite{kashdan, final}. Thus, integrating these and similar features to the age, gender, and socioeconomic status which are currently integrated into the framework, might help to obtain even better outcomes. 

\section*{Declarations}
\subsection*{Funding}
This research did not receive any specific grant from funding agencies in the public, commercial, or not-for-profit sectors.

\subsection*{Conflicts of interest/Competing interests}
None.

\subsection*{Data availability}
The data that has been used in this study is available by a formal request from the authors. 

 
\bibliography{biblio}
\bibliographystyle{unsrt}

\end{document}